\documentclass[PRA,showpacs,twocolumn]{revtex4}
\usepackage{amssymb}
\usepackage{epsfig}
\usepackage{graphicx}
\usepackage{color}
\begin{document}
\title{Displaced dynamics of binary mixtures  in linear and nonlinear optical lattices}
\author{Golam Ali Sekh, Mario Salerno}
\affiliation{Dipartimento di Fisica "E. R. Caianiello", via ponte don Melillo I-84084, Fisciano (SA), Italy}
\author{ Aparna Saha, and  Benoy Talukdar}
\affiliation{Department of Physics, Visva-Bharati University, Santiniketan 731 235, India}
\date{\today}
\begin{abstract}
The dynamical behavior of matter wave solitons of two-component Bose-Einstein condensates (BEC) in combined linear and nonlinear optical lattices (OLs) is investigated. In particular, the dependence of  the frequency of the oscillating dynamics resulting from initially slightly displaced components is investigated  both analytically, by means of a variational  effective potential approach for the reduced collective coordinate dynamics  of the soliton, and numerically,  by direct integrations of the mean field equations of the BEC mixture.  We show that for small initial displacements binary solitons can be viewed as point masses connected by elastic springs of strengths related to  the amplitude  of the OL and to the intra and inter-species interactions. Analytical  expressions of  symmetric and anti-symmetric mode frequencies, are derived and occurrence of beatings phenomena in the displaced dynamics is predicted. These  expressions are shown to  give a very good estimation of the oscillation frequencies for different values of the intra-species interatomic scattering length, as confirmed by direct  numerical integrations of the mean field Gross-Pitaevskii equations (GPE) of the mixture.
The possibility to use displaced dynamics for  indirect  measurements of BEC mixture characteristics such as  number of atoms and interatomic interactions is also suggested.
\end{abstract}
\pacs{03.75.-b,67.85.Hj, 05.45.Yv}
\maketitle

\section{Introduction}
Binary mixtures of  Bose-Einstein condensates (BECs) are presently attracting a great deal of interest  in connection with a series of interesting phenomena such as the formation of segregate domains \cite{1}, polarized states \cite{2}, spin textures \cite{3}, topological excitations \cite{4}, novel Josephson oscillations \cite{5,6-Mazzarella-2009} Rabi – Josephson oscillations \cite{7-Rabi-Josephson}, four wave mixing \cite{8}, etc. Moreover, multi-component BECs have been  shown to support nonlinear waves of novel type such as symbiotic solitons \cite{9}, domain-wall solitons \cite{10} and combinations of dark-dark \cite{11} and  bright-dark solitons \cite{12,13}, the last one  leading to long lived oscillations which were experimentally observed in \cite{14}.
The possibility to trap binary mixtures in optical lattices (OLs), experimentally demonstrated in \cite{15}, has added further interest to the field. In particular, the interplay between the nonlinearity induced by the interatomic interactions and the strength of  the  OL has been shown to lead to interesting phenomena such as Landau-Zener tunneling  \cite{16}, transitions from superfluids to Mott insulators \cite{17}, Moreover, existence of nonlinear periodic waves on nonzero backgrounds \cite{18-kostov2004},   gap solitons \cite{19},  mixed-symmetry modes and breathers both in continuous and discrete (arrays) mixtures  \cite{20-cruz2007}.

Besides usual (e.g. linear) OLs,  it is also possible to introduce  a periodic structure in the system by  modulating the scattering lengths in space by  means of the Feshbach resonance technique \cite{21}. This allows to create what is known as a {\it nonlinear optical lattice} (NOL). BEC mixtures in NOLs have been recently considered in connection with quantum simulation of novel Hubbard models \cite{22} and interesting phenomena such as sonic analogues of black holes \cite{23} and  control of soliton creation  \cite{24}. A possibility of observing delocalizing transition even in one-dimensional BECs loaded in OLs due to the presence of the NOL has been also suggested \cite{25}. For a fresh review on BECs in nonlinear optical lattices we refer the article in \cite{26}.
 In all these studies, however, the effects of a combined linear and nonlinear optical lattice  on the  soliton dynamics and  the link between dynamical behaviors and interactions,  have been scarcely investigated.

The aim of the present paper is to study the mean field  dynamics of initially displaced soliton components  of  binary BEC  mixtures in the presence of a combined linear and nonlinear OL.
In particular, the dependence of  the frequency of the resulting oscillating dynamics
on the inter-species interaction and on the number of atoms is investigated. This is done  both analytically, by means of a variational  effective potential for the displaced dynamics, and numerically  by direct integrations of the mean field equations of the BEC mixture. We show that in the limit of small initial displacements, the effective potential leads to a mechanical interpretation of  a binary soliton motion in terms of two point masses connected by elastic springs of strengths related to OL's amplitude and  to the intra and inter-species interactions. The displaced dynamics, being  the same as the one  of coupled harmonic oscillators, can be decomposed in term of a normal mode analysis from which analytical  expressions of the symmetric and anti-symmetric mode frequencies, are explicitly derived. These  expressions are shown to  give a very good estimation of the oscillation frequencies for different values of the intra-species interatomic scattering length, as confirmed by direct  numerical integrations of the mean field Gross-Pitaevskii equations (GPE) of the mixture. The occurrence of beating phenomena for unequal and for equal numbers of atoms in the mixture  for small interspecies interactions, is also discussed. The stabilities of stationary  and oscillating dynamics are investigated by Vakhitov-Kolokolov (VK) criterion \cite{27} and by numerical simulations, respectively. These results suggest  the possibility to use dynamical behaviors of suitably prepared initial multi-component BEC solitons  as a tool for extracting  information about  physical characteristics of BEC mixtures such as interatomic interactions and species populations.

The paper is organized as follows. In Sec. II we introduce the mean field model equations describing BEC mixtures in combined linear and nonlinear optical lattices and derive a  variational effective potential formulation  for the matter waves soliton dynamics.In section III we consider the displaced binary soliton dynamics in the framework of a coupled harmonic oscillator model which is  valid in the limit of small initial displacements. Analytical expressions for the symmetric and antisymmetric mode frequencies are explicitly derived. In Sec. IV results of displaced soliton dynamics obtained by direct numerical integrations of the GPE are compared  with the analytical predictions. The stability of stationary two component solitons and their slightly displaced dynamics are also investigated. Finally, in Sec. V the main results of the paper  are briefly summarized.

\section{Model equation and variational analysis}
 We consider as a mean field model for a mixture of two homonuclear condensates \cite{28} in an external trapping potential, the following system of coupled Gross-Pitaevskii equations
\begin{eqnarray}
i\hbar \frac{\partial \phi_j}{\partial t}&=&-\frac{\hbar^2}{2m}\frac{\partial \phi_j}{\partial x^2}+V_{ext}(x) \phi_j +2 \hbar \omega_{\perp} a_{s}^{(1)}|\phi_j|^2
\phi_j\nonumber\\&+&2\hbar\omega_\perp a_s^{(12)}|\phi_{3-j}|^2\phi_j.
\label{eq1}
\end{eqnarray}
where  $\phi_j$ $(j=1,2)$ denote the wave function of the binary  mixture and  $V_{ext}(x)$ the external potential resulting from harmonic and optical lattice confinement, in the following taken of the form
\begin{eqnarray}
V_{ext}(x)=\frac{1}{2} m w_x^2 x^2 +V_L \cos(2 k_L x).
\label{eq2}
\end{eqnarray}
Here $\omega_x$ and $\omega_\perp$ are the longitudinal and transverse trapping frequencies of the harmonic confinement,  $a_s^{(1)}$ and  $a_s^{(12)}$ are the intra- and inter-species scattering lengths, $V_L$ and $k_L$ are respectively strength and wave number of the optical lattice. Since the longitudinal harmonic confinement introduces only slight modifications to the longitudinal periodic potential (in experimental settings $\omega_x$ is of the order of a few Hz), it will  be ignored in the following \cite{29}. Introducing dimensionless variables:
\begin{eqnarray}
\tau=t \frac{\hbar}{E_r},\,\,E_r=\frac{\hbar^2 k_L^2}{2 m},\,\,\,s=x k_L\,\,\,{\rm and}\,\,\,\psi_j=\frac{\phi_j}{\sqrt{k_L}}
\label{eq3}
\end{eqnarray}
Eqs. \ref{eq1} can be written in the form
\begin{eqnarray}
i \frac{\partial \psi_j}{\partial \tau}&=&-\frac{1}{2}\frac{\partial \psi_j}{\partial s^2}+V_0 \cos(2 s) \psi_j +g_{11}|\psi_j|^2
\psi_j\nonumber\\&+&g_{12}|\psi_{3-j}|^2\psi_j
\label{eq5}
\end{eqnarray}
where $V_{0}=\frac{V_L}{E_r}$ and $g_{11}=2a_s^{(1)}k_L$ and $g_{12}=2a_s^{(12)}k_L$ are rescaled intra- and inter-species interaction strengths.
In  this Eqs. the order parameter $\psi_j$ is normalized to the total number of atoms such that
$ \int_{-\infty}^{+\infty}\left(|\psi_1|^2+|\psi_2|^2\right) ds = N_1 + N_2$, where $N_j, \, j=1,2$ are the separately conserved numbers of atoms in each component.
In the following we fix $k_L=2$ and assume a dependence of  the intra-species interaction of the form
\begin{equation}
g_{11}=g_{11}^{(0)} +g_{11}^{(1)}\cos(2 s)
\label{NOL}
\end{equation}
with the spatial modulation part denoting a NOL of strength $g_{11}^{(1)}$. In an  experimental context such a spatial modulation could  be produced by optically induced Feshbach resonances \cite{OFR2} e.g. by a
laser field  tuned near a photo association transition. Virtual radiative
transitions of a pair of interacting atoms to this level can then
change the value and even  reverse the sign of the scattering
length. It can be  shown that a modulation of the laser field intensity of the form $I=I_0 \cos^2(\kappa x)$ reflects in a modulation of the scattering length of the form $a_s^{(1)}(x) = a_{s0}^{(1)}[1 + \alpha I/(\delta +I)]$,
where $a_{s0}^{(1)}$ is the intra-species scattering length in the absence of light, $\delta$ is
the frequency detuning of the light from the resonance, and $\alpha$ is a
constant factor \cite{OFR2,SM}. For weak intensities $I_0 \ll |\delta|$ the real part of the scattering length can be then approximated as
$a_s^{(1)}=a_{s0}^{(1)}+ a_{s1}^{(1)} \cos^2(\kappa x)$ which is
essentially the same form  assumed in Eq.~(\ref{NOL}).

Note that in the absence of the OLs and  with $g_{12}=0$, Eq. (\ref{eq5}) decouple into two nonlinear Schr\"odinger equations  which admit, for  attractive
intra-species interactions, exact bright soliton solutions with typical Gaussian-like function  shape. With the view  to solve Eqs. in (\ref{eq5}) within a variational approach, we then adopt  for the  coupled soliton wavefunction the following ansatz
\begin{eqnarray}
\psi_j(s,\tau)&=&A_j \exp[-\frac{(s-s_{0j})^2}{2 a_j^2}+i(\dot{s}_{0j}(s-s_{0j}) + \Phi_j)],\nonumber\\&&\;\;\;\;\;j=1,2
\label{eq6}
\end{eqnarray}
with parameters $A_j$, $a_j$,  $s_{0j}$,  $\Phi_j$, denoting  amplitude, width,  center of mass and phase of the soliton,  respectively, taken in the following as time-dependent parameters. Note that the wavefunction is normalized to the total number of atoms $N_j$ so that $A_j=\sqrt{\frac{N_j}{\sqrt{\pi} a_j}}$.

The effective Lagrangian for the system is written as
$\left\langle{\cal L}\right\rangle=\int_{-\infty}^{\infty} L ds$ with the Lagrangian density $L$ given by
\begin{eqnarray}
<L>&=&\sum_{j=1}^2\sqrt{\pi}a_jA_j^2\left[\frac{1}{4 a_j^2}+\frac{g_{11}^{(0)}}{2\sqrt{2}}  A_j^2+ V_0 e^{-a_j^2} \cos(2 s_{0j})\right.\nonumber\\&+&\left.\frac{g_{11}^{(1)}}{2\sqrt{2}} e^{- a_j^2/2} A_j^2 \cos(2 s_{0j})-\frac{1}{2} \dot{s}_{0j}^2+  \dot{\Phi}_j\right]\nonumber\\&+&g_{12} a_1a_2A_1^2A_2^2
\frac{\exp[-\frac{(s_{01}-s_{02})^2}{a_1^2+a_2^2}]}
{\sqrt{a_1^2+a_2^2}}.
\label{eq7}
\end{eqnarray}
From the Ritz optimization conditions \cite{34}, we have $\frac{\delta\langle{\cal L}\rangle }{\delta \Phi_{j}}=0$, $\frac{\delta\langle{\cal L}\rangle }{\delta A_{j}}=0$,$\frac{\delta\langle{\cal L}\rangle }{\delta a_{j}}=0$
and  $\frac{\delta\langle{\cal L}\rangle }{\delta s_{oj}}=0$. The first optimization condition
\begin{eqnarray}
\frac{d}{d\tau}\left[\sqrt{\pi}a_jA_j^2\right]=0,
\label{eq8}
\end{eqnarray}
in conjunction with the normalization condition of $\psi_j$ implies that $\sqrt{\pi}a_jA_j^2=N_j$ is a constant. 
This constrain when used in the relations obtained from the other optimization conditions give
\begin{eqnarray}
&&\frac{1}{2 a_j^2}+
g_{11}^{(0)} \sqrt{\frac{2}{\pi}} \frac{N_j}{a_j}+2 V_0 e^{-a_j^2}  \cos(2 s_{0j})-\dot{s}_{0j}^2+2\dot{\Phi}_j\nonumber\\&+&g_{11}^{(1)}\sqrt{\frac{2 }{\pi}} N_je^{-\frac{ a_j^2}{2}}   \cos(2 s_{0j})
\nonumber\\&+&\frac{2g_{12}}{\sqrt{\pi}} N_{3-j}
\frac{\exp[-\frac{(s_{01}-s_{02})^2}{a_1^2+a_2^2}]}
{\sqrt{a_1^2+a_2^2}} =0,
\label{eq9}
\end{eqnarray}
\begin{eqnarray}
&-&\frac{1}{2 a_j^3} + \frac{g_{11}^{(0)}}{\sqrt{2\pi}} \frac{N_j}{a_j^2}+ \frac{2V_0}{a_j} (1-2 a_j^2) e^{-a_j^2} \cos(2 s_{0j})\nonumber\\&+&\frac{g_{11}^{(1)}N_j}{\sqrt{2 \pi}a_j^2}(1-a_j^2)e^{-a_j^2/2}
\cos(2 s_{0j})-\frac{\dot{s}_{01}^2}{a_1^2}+\frac{2\dot{\Phi}_1}{a_j}\nonumber\\&+&
\frac{g_{12} N_{3-j}}{\sqrt{\pi}a_j}\left[a_{3-j}^4+a_j^2a_{3-j}^2+2a_j^2(s_{01}-s_{02})^2
\right]\nonumber\\&& \times\frac{e^{-\frac{(s_{01}-s_{02})^2}{a_1^2+a_2^2}}}{(a_1^2+a_2^2)^{5/2}}
\label{eq10}
\end{eqnarray}
and
\begin{eqnarray}
&&\ddot{s}_{0j}-2  V_0 e^{-a_j^2} \sin(2 s_{0j})-\frac{g_{11}^{(1)} N_j e^{- a_j^2/2}}
{\sqrt{2 \pi} a_j}\sin(2 s_{0j})\nonumber\\&+&(-1)^j\frac{2 g_{12}N_{3-j} (s_{01}-s_{02})}{\sqrt{\pi} (a_1^2+a_2^2)^{3/2}} e^{-\frac{(s_{01}-s_{02})^2}{a_1^2+a_2^2}}=0.
\label{eq11}
\end{eqnarray}
\begin{figure}[h]
\includegraphics[width=6cm, height=4.0cm]{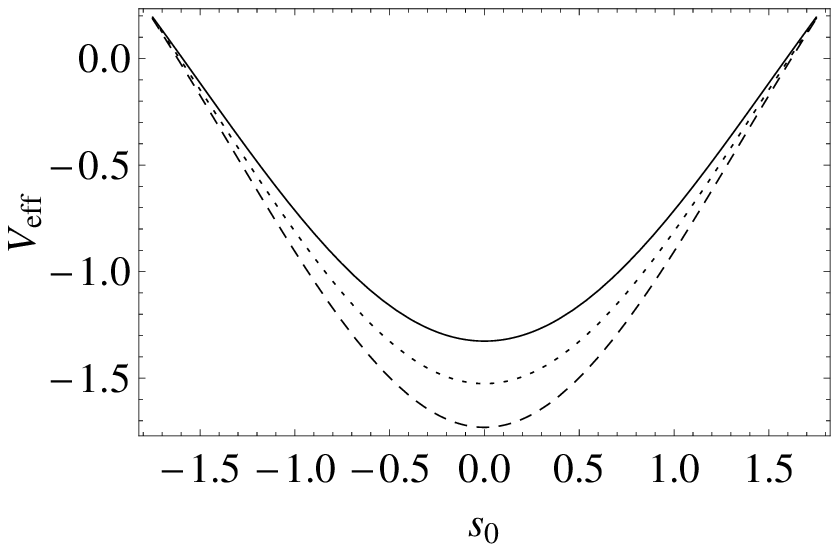}
\includegraphics[width=6cm, height=4.0cm]{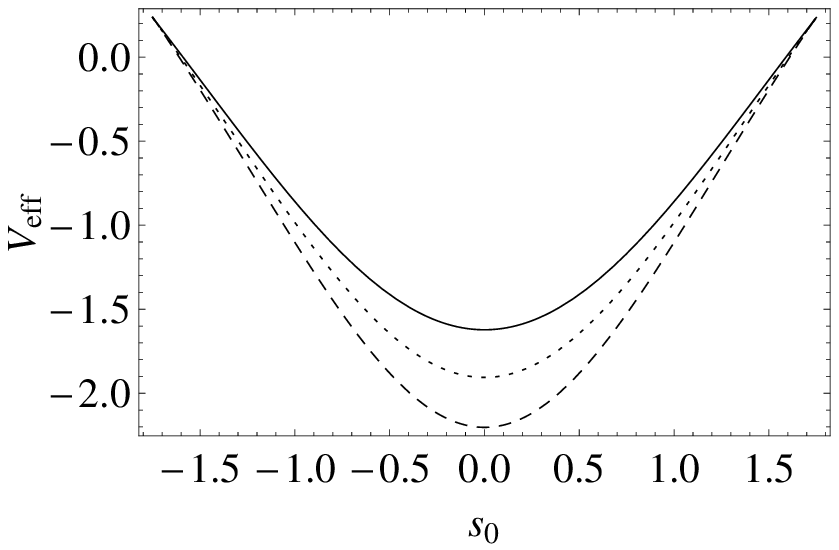}
\includegraphics[width=6cm, height=4.0cm]{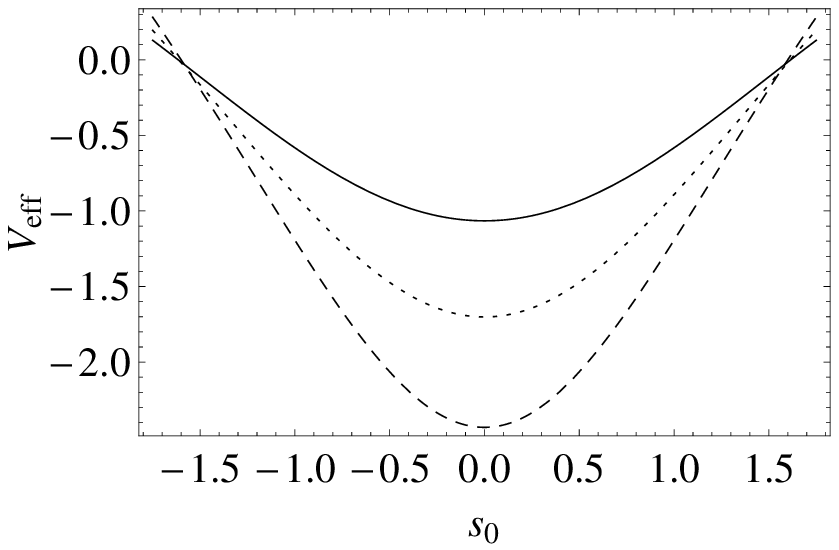}
\caption{Effective potential versus separation for  $ V_0 = -0.$5. Top panel gives $V_{\rm eff}$ with $N_1=1.0$ and $N_2=0.5$ for different values of $g_{12}$, namely, $-0.2$ (solid), $-0.4$ (dotted) and $-0.6$(dashed). The middle panel gives $V_{\rm eff}$ with $N_1=1.0$ and $N_2=1.0$ for different values of $g_{12}$, namely, $-0.2$ (solid), $-0.4$ (dotted) and $-0.6$(dashed). The bottom panel shows
$V_{\rm eff}$ with $g_{12}=-0.5$ for different values of $N=N_1=N_2$, namely, $N=0.4$ (solid), $0.8$ (dotted) and $1.2$  (dashed).}
\label{fig1}
\end{figure}

In order to derive an explicit formula for the effective interacting potential of the coupled solitons, we consider that the condensates are symmetrically placed with respect to the a OL minimum i.e, $s_{0j}=\pm s_0/2$. In this case, Eqs. in (\ref{eq11}) can be combined to give
\begin{eqnarray}
 && \ddot{s}_0 -2  V_0 \left( e^{-a_1^2} +  e^{-a_2^2}\right)\sin( s_0)\nonumber\\&-&\frac{ 2 g_{12} (N_1+ N_2) s_0}{\sqrt{\pi} (a_1^2+ a_2^2)^{3/2}}e^{-\frac{s_0^2}{a_1^2+a_2^2}}\nonumber\\ &-&
\frac{ g_{11}^{(1)}}{\sqrt{2\pi}}\left(\frac{N_1}{ a_1}e^{-a_1^2/2} +\frac{N_2}{ a_2}e^{-a_2^2/2}\right)\sin(s_0)=0
\label{eq13}
\end{eqnarray}
as the  evolution equation for the separation $s_0$ between center of the solitons.
Notice that Eq. (\ref{eq13}) is the same as the dynamics of a  Newtonian particle in the effective potential
\begin{eqnarray}
V_{\rm eff}(s_0)&=&\left[ 2  V_0 \left(e^{-a_1^2} + e^{-a_2^2}\right)\cos( s_0)\right.\nonumber\\&+&\left.\frac{ g_{11}^{(1)}}{\sqrt{2\pi}}\left(\frac{N_1}{ a_1}e^{-a_1^2/2} +\frac{N_2}{ a_2}e^{-a_2^2/2}\right)\cos(s_0)\right.\nonumber\\&+&\left.\frac{  g_{12} (N_1+ N_2) }{\sqrt{\pi} (a_1^2+ a_2^2)^{1/2}} e^{-\frac{s_0^2}{a_1^2+a_2^2}}\right].
\label{eq14}
\end{eqnarray}
Also notice that this potential  has the absolute minimum in the origin and that for small values of $s_0$ around the minimum of the potential can  be approximated as a harmonic potential. In such approximations, the small oscillation frequency of displaced solitons dynamics can be written as
\begin{eqnarray}
\omega&=&\left[-2 V_0( e^{- a_1^2}+ e^{-a_2^2}) -\frac{2 g_{12} (N_1+ N_2)}{(a_1^2+a_2^2)^{3/2} \sqrt{\pi}}\right.\nonumber\\ &-&\left.\frac{g_{11}^{(1)}}{\sqrt{2\pi}}\left(\frac{N_1}{a_1 }e^{-a_1^2/2}+\frac{N_2}{a_2}e^{-a_2^2/2}\right)\right]^{1/2}.
\label{eq15}
\end{eqnarray}

Moreover, one can show that the vanishing condition of $\frac{\delta\langle{\cal L}\rangle }{\delta A_{j}}$ gives the chemical potential, $\mu$,  of stationary components as:
\begin{eqnarray}
\mu_j&=&\frac{1}{4 a_j^2}+ \frac{g_{11}^{(0)} N_j}
{a_j \sqrt{2 \pi}} +\frac{g_{11}^{(1)}N_j}{a_j \sqrt{2\pi}} e^{-a_j^2/2}  \cos(s_0)\nonumber\\&+&  V_0 e^{-a_j^2} \cos(s_0) +\frac{g_{12}N_{3-j}}{\sqrt{\pi}}\frac{e^{-\frac{s_0^2}{a_1^2 + a_2^2}}}{\sqrt{a_1^2 + a_2^2}}
\label{eq16}
\end{eqnarray}
(in writing  Eq. (\ref{eq16}) we have used  $\Phi_j=-\mu_j \tau$ and $\dot{s}_{0j}=0$ in Eq. (\ref{eq9})). This expression can be used (see below) to study the stability of stationary two component solitons through the  Vakhitov-Kolokolov (VK) criterion.
From Eq. (\ref{eq14}) we see that the effective potential  for the coupled solitons dynamics is highly anharmonic and  consists of  three terms: the first two arise from the linear and nonlinear optical lattices while the third one comes from the mutual interaction between the solitons.
The mutual interaction term depends both on the number of atoms in the condensates and on the strength of the interactions. This  part of the potential will therefore change sensitively with the variation of $N$ and $g_{12}$.

In Fig. \ref{fig1} we show the effective potential as a function of $s_0$  for two attractively interacting solitons and different values of  $-g_{12}$ (left and middle panels) and $N=N_1=N_2$(right panel). More specifically, left panel give $V_{eff}$ with $N_1=1$ and $ N_2=0.5$ while middle panel shows $V_{eff}$ with $N_1=N_2=1$ for different values of $-g_{12}$.
Note that the inter-species interaction is effective mainly for BEC components with a significant spatial  overlapping e.g. when they are very close to each other. In this situation an oscillatory dynamics of the BEC components around their common center of mass can  be started  by slightly  displacing them  from the equilibrium position corresponding to the fundamental  minimum  of the effective potential in Fig. \ref{fig1}. Also note that for an attractive inter-species interaction the absolute minimum of the effective potential becomes deeper and deeper as $N |g_{12}|$ is increased. Therefore, the reduced equation of motion in (\ref{eq15})  implies that the solitons oscillation with respect to each other if they are placed very close to the effective potential minimum at $s_0=0$.
For repulsive  inter-species interactions, however, the effective potential will have the shape of a  barrier (rather than a potential well) with a maximum (rather than a minimum)  at the origin. In this case, the soliton components  move away from each other keeping their  shapes unchanged \cite{32}.
\begin{figure}[h]
\includegraphics[width=6cm,height=3.5cm]{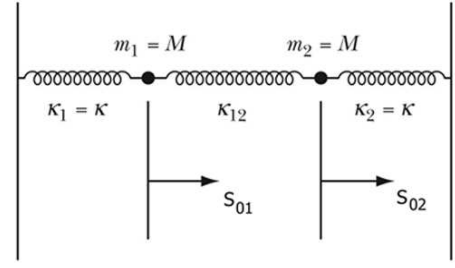}
\caption{Mechanical model of displaced binary soliton dynamics in terms of harmonic oscillators of elastic constant  $\kappa$ coupled  by a spring of elastic constant $\kappa_{12}$. }
\label{fig2}
\end{figure}

\section{Normal mode analysis of displaced binary soliton dynamics}
It is useful to gain some modeling insight of the displaced binary soliton dynamics in the limit of small displacements  $s_{01}\ll 1$ and $s_{02}\ll 1$. In this case Eq. (\ref{eq11}) reduces to
\begin{eqnarray}
\ddot{s}_{0j}&-&\left(4V_0 e^{-a_j^2}+\frac{2g_{11}^{(1)}N_j}{\sqrt{2\pi}a_j}e^{-a_j^2/2}+\frac{2 g_{12} N_{3-j}}{\sqrt{\pi}(a_1^2+a_2^2)^{3/2}}\right)s_{0j}\nonumber\\&+&\frac{2 g_{12} N_{3-j}}{\sqrt{\pi}(a_1^2+a_2^2)^{3/2}}s_{0{3-j}}=0.
\label{eq12}
\end{eqnarray}
Let us concentrate  for simplicity on binary solitons with equal number of atoms and equal widths e.g. $N_1=N_2=N$ and $a_1=a_2=a$. Introducing parameters
\begin{eqnarray}
&& M=\frac{\sqrt{2 \pi} a^3}{N}, \;\;\;\;\;\; \kappa_{12}=- g_{12},\; \\ &&
\kappa = - 4 V_0 \frac{\sqrt{2 \pi} a^3 e^{- a^2}}{N}  - 2 g^{(1)}_{11} a^2 e^{- a^2/2},
\end{eqnarray}
we can rewrite   Eq.  (\ref{eq12})  in the form
\begin{eqnarray}
\ddot{s}_{01}&=& - \frac{\kappa + \kappa_{12}}{M} s_{01} + \frac{\kappa_{12}}{M} s_{02}, \\
\ddot{s}_{02}&=& - \frac{\kappa + \kappa_{12}}{M} s_{02} + \frac{\kappa_{12}}{M} s_{01}, \\
\label{coupledha}
\end{eqnarray}
which are the same as the equation of motion of two coupled identical harmonic oscillators of mass $M$ and  elastic spring $\kappa$ connected by a spring of elastic constant $\kappa_{12}$ (see Fig. \ref{fig2}). In the absence of inter-species interaction, (as it is the case, for example,  when the interspecies scattering length is detuned to zero by means of a Feshbach resonance)  In the presence of interspecies interactions the above equations are readily decoupled in the normal mode coordinates: $\xi_1=s_{01}-s_{02}$, $\xi_2=s_{01}+s_{02}$, this giving $M \ddot{\xi_i}= -\omega_i^2 \xi_i$, $i=1,2$, with characteristic frequencies
\begin{equation}
\omega_1=\pm \sqrt{\frac{\kappa + 2 \kappa_{12}}{M}}, \;\; \omega_2=\pm \sqrt{\frac{\kappa}{M}}
\end{equation}
and explicit normal mode  solutions
\begin{equation}
\xi_i (t)=A_i^{+} e^{i\omega_i t} + A_i^{-} e^{- i\omega_i t},\;\; \;\; i=1,2.
\label{nmod}
\end{equation}
The most general solution of the displaced soliton dynamics in the coupled harmonic oscillator approximation follows  from Eq. (\ref{nmod}) as
\begin{equation}
s_{01}(t)=\frac {1}{2} (\xi_2(t)+\xi_1(t)), \;\; s_{02}(t)=\frac{ 1}{2} (\xi_2(t)-\xi_1(t)).
\label{gsolution}
\end{equation}
From these equations we see that  the solution $\xi_1$ associated to the frequency $\omega_1 \equiv \omega_{asym}  $ corresponds to an asymmetric (out of phase)  oscillation of the displaced two component soliton, while the solution $\xi_2$ corresponds to a  symmetric (in phase) motion of frequency $\omega_2 \equiv \omega_{sym}$ in which the  coupling spring remains unstretched. Notice that $\omega_{asym}$ is the same as the expression
of the frequency derived in Eq. (\ref{eq15}). Also note that in analogy with optical and acustical vibrations of molecules, this frequency, for attractive inter- and intra-species interactions,  is always higher than the frequency $\omega_{sym}$ of the symmetric mode, e.g.   $\omega_{asym}/ \omega_{sym} \ge  1$, with the equality holding in the case $g_{12}=0$. More explicitly, the following dependence  for the frequency ratio of asymmetric and symmetric modes on parameters of the binary mixture,
is derived:
\begin{eqnarray}
\nu_r&\equiv&\frac{\omega_{asym}}{\omega_{sym}}\nonumber\\
&=&[1+ \frac{N g_{12} e^{a^2/2}/a^2}{2 V_0 \sqrt{2 \pi} a e^{-a^2/2} +  N g_{11}^{(1)}}]^{1/2}.
\label{eqrfreq}
\end{eqnarray}
Note that in the weak coupling limit $|g_{12}| \ll 1$ the displaced dynamics will display typical
beating phenomena with a
high frequency component oscillating  inside a slowly varying envelope, with beating frequencies
$\omega_{beat}=\omega_{asym}-\omega_{sym}$,  plus  order combinations.

For the general case   $N_1\neq N_2$ (equivalently, $a_1\neq a_2$), the dependence of characteristic
frequencies of  the oscillators on parameters can be derived in  similar manner.  In the next section
we shall compare the above predictions  for the soliton displaced  dynamics with direct numerical integrations
of the coupled GPE in (\ref{eq5}).

%
\begin{figure}[h]
\includegraphics[width=6cm,height=3.5cm]{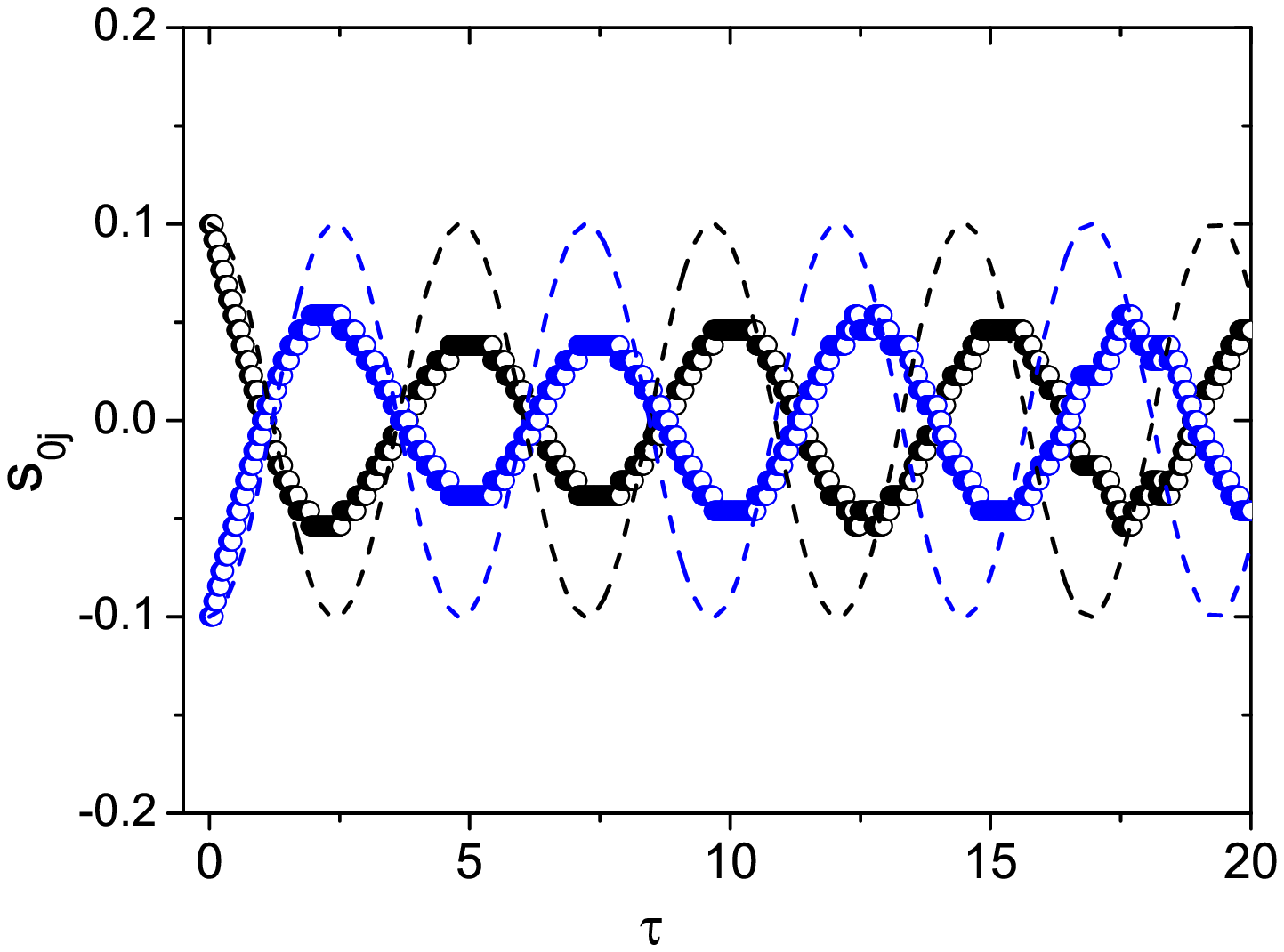}
\includegraphics[width=6cm,height=3.5cm]{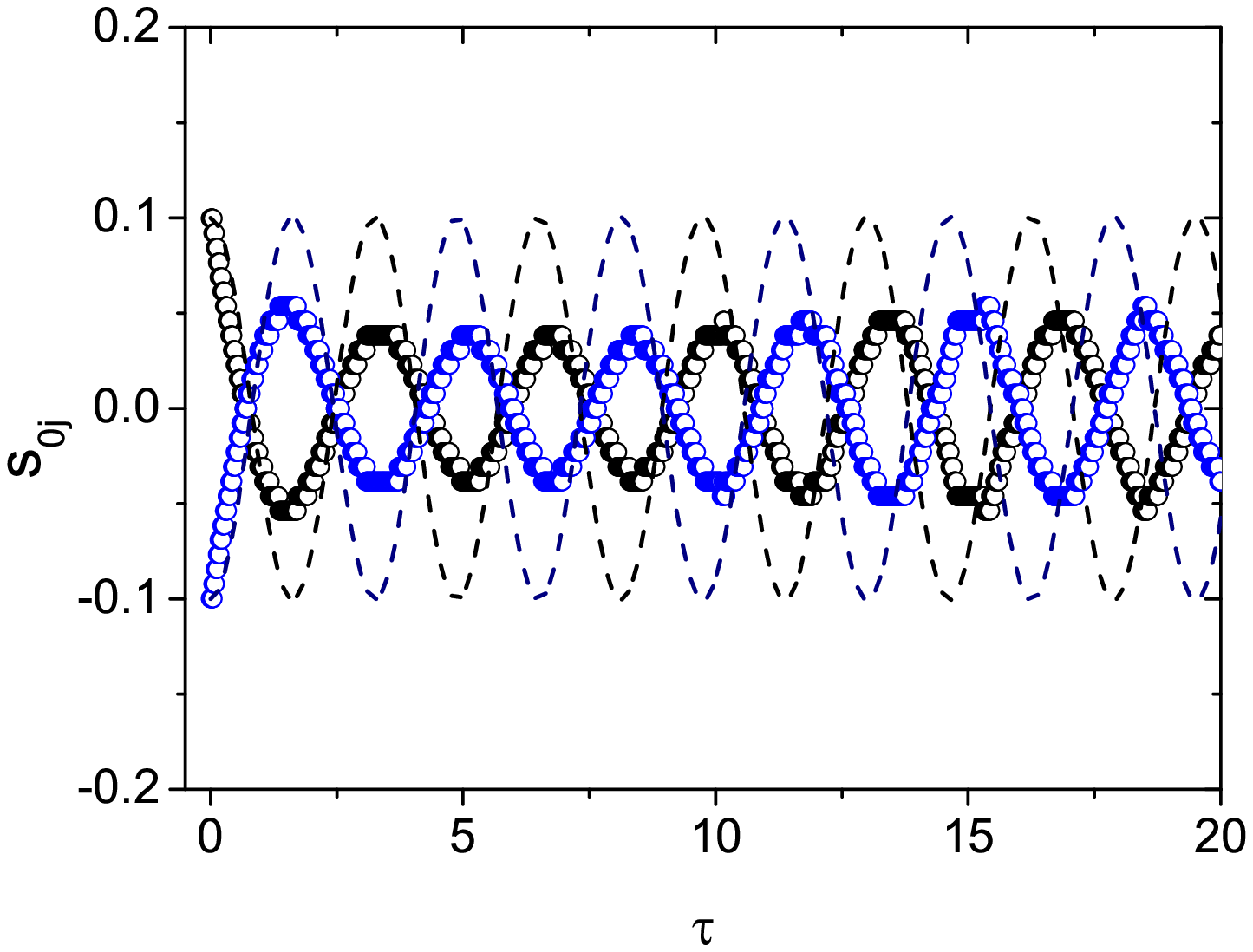}
\includegraphics[width=6cm,height=3.5cm]{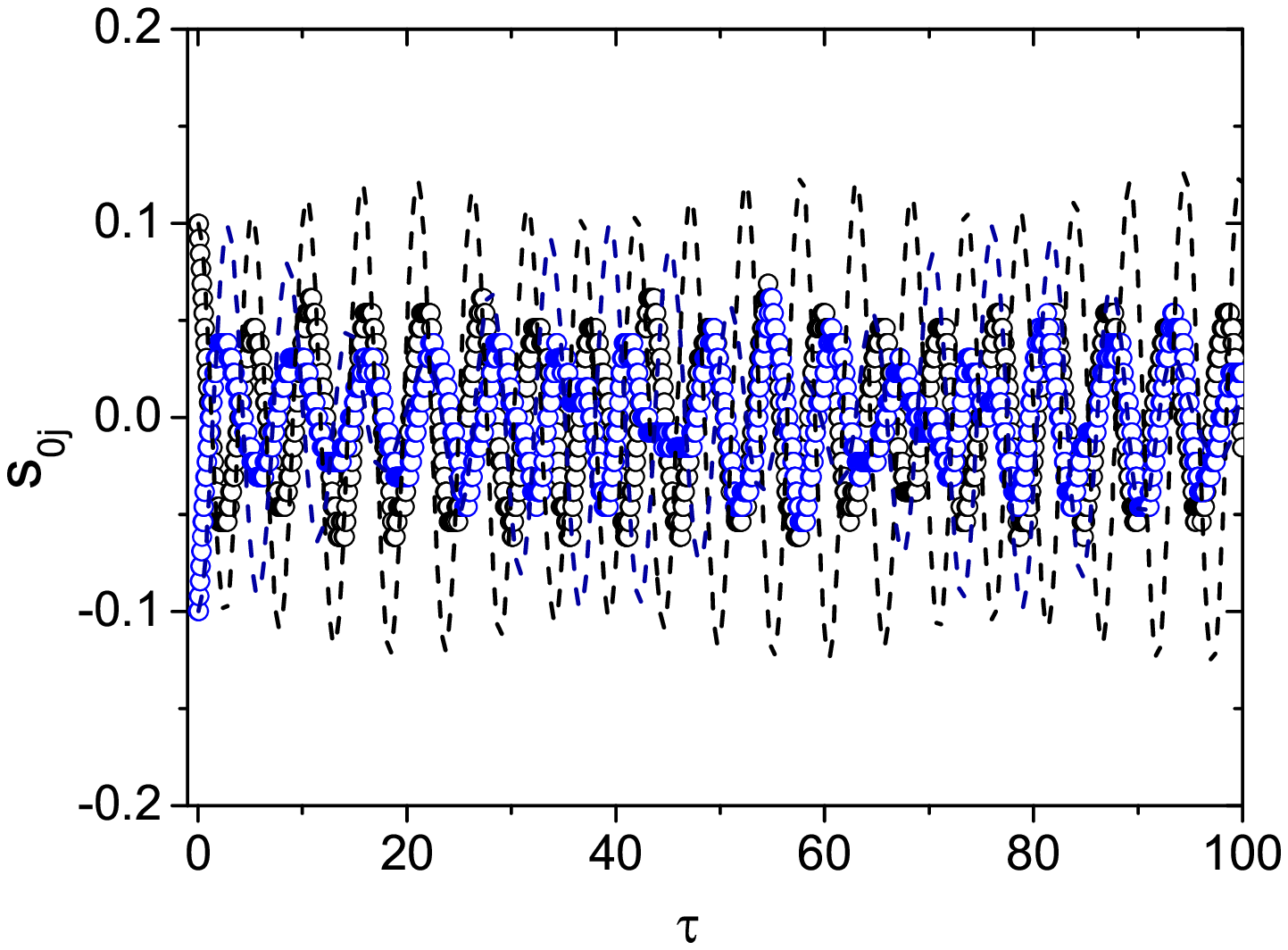}
\includegraphics[width=6cm,height=3.5cm]{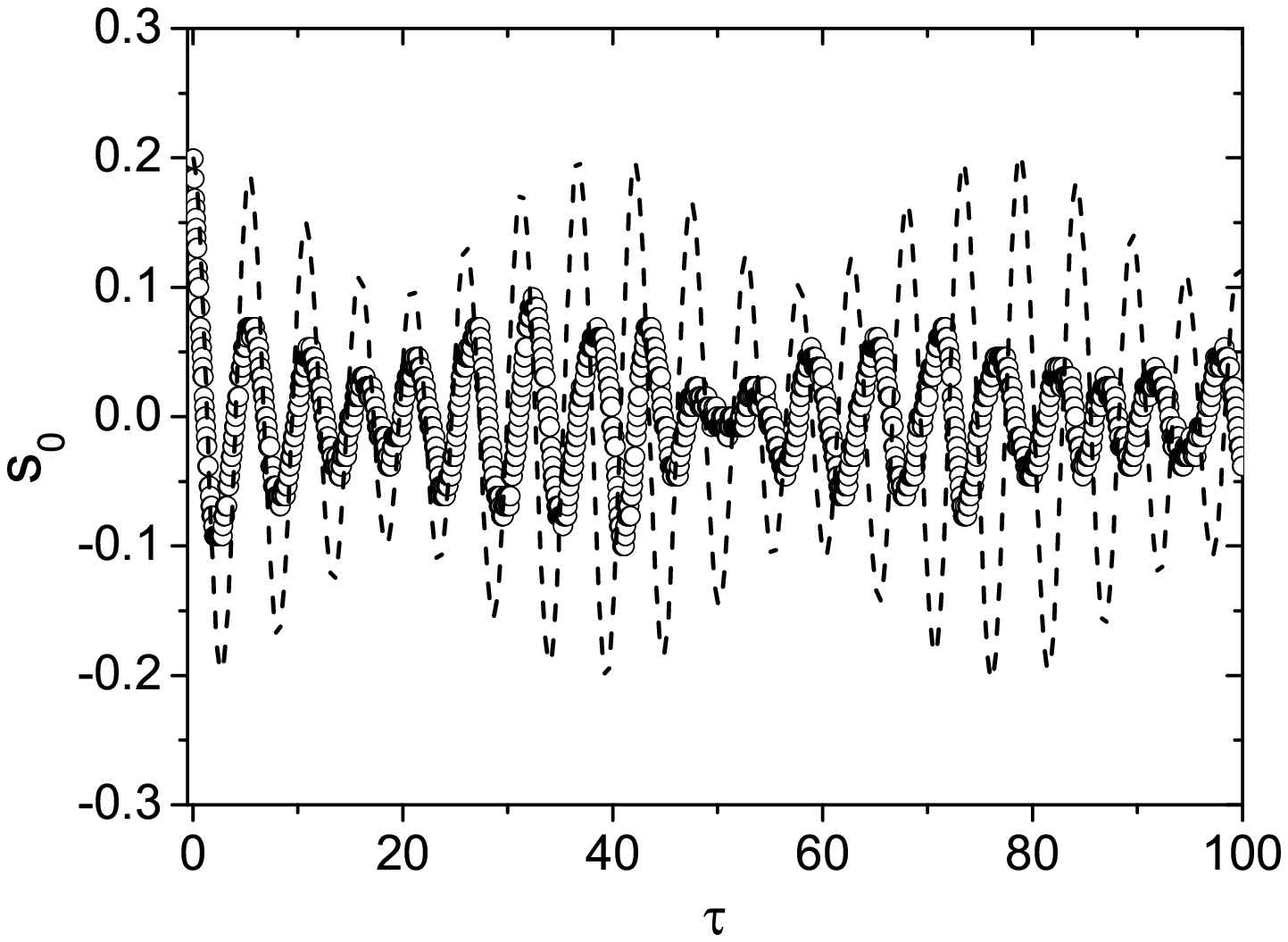}
\vskip -0.5cm
\caption{Oscillation of coupled BEC components  for $ V_0 = -0.5,\, g_{11}^{(0)} = -1,\, g_{11}^{(1)} = -0.5$ and $s_{0}=0.2$. Here, number of panels is counted from the top. First panel shows motion of soliton profiles for $N_1=N_2=1.0$ and $g_{12}=-0.2$. Second panel. Same as in first panel but for $N_1=N_2=1.4$ and $g_{12}=-0.5$. Third panel. Motion of the centers of soliton components  for unequal number of atoms $N_1=1$, $N_2=0.5$ anf for  $g_{12}=-0.2$. The bottom panel show beatings arising from  the superposition of the oscillatory components displayed in the 3rd panel. In all panels curves with big circles give results for PDEs in   (\ref{eq5})  while dashed curves represent results for ODEs in (\ref{eq11}).}
\label{fig3}
\end{figure}

\section{Dynamics and stability of displaced binary solitons: Numerical results }
In the following coupled GPE numerical investigations we assume  that solitons prepared in such a manner that their relative coordinates  are located at  small distances  from the minimum of $V_{eff}$ in Fig. \ref{fig1}. We remark that initial small displacements of the two components of the mixture could be experimentally induced by a rapid change of the inter-species scattering length from negative to positive and then to negative again, by means of the Feshbach resonance technique. The inversion of the sign of the interaction for a small fraction of time can be achieved with a properly designed  time-dependent external  magnetic field. The component solitons will move in  opposite directions during the  short repulsive inter-species interaction time, and will become  slightly separated (separation can be made small by properly reducing the repulsive time). Taking into account  detectable length scales of real experiments\cite{35}, we use  in most of the calculations $s_0=0.2$, although larger initial  displacements ($s_0 \approx 1$) will also be used for anharmonic effects.

\begin{figure}[htb]
\centerline{
 \includegraphics[width=4.5cm,height=4cm]{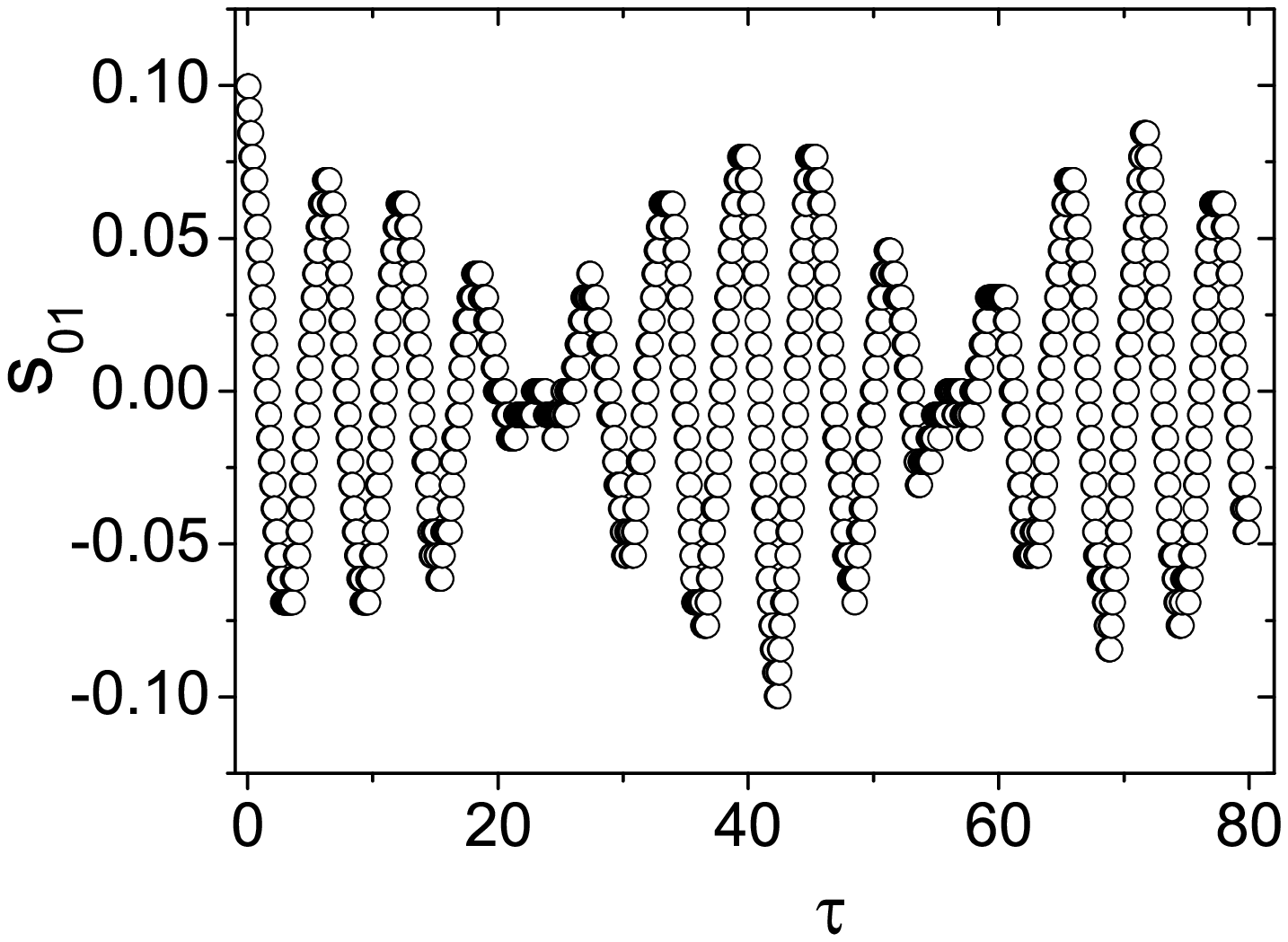}
\includegraphics[width=4.5cm,height=4cm]{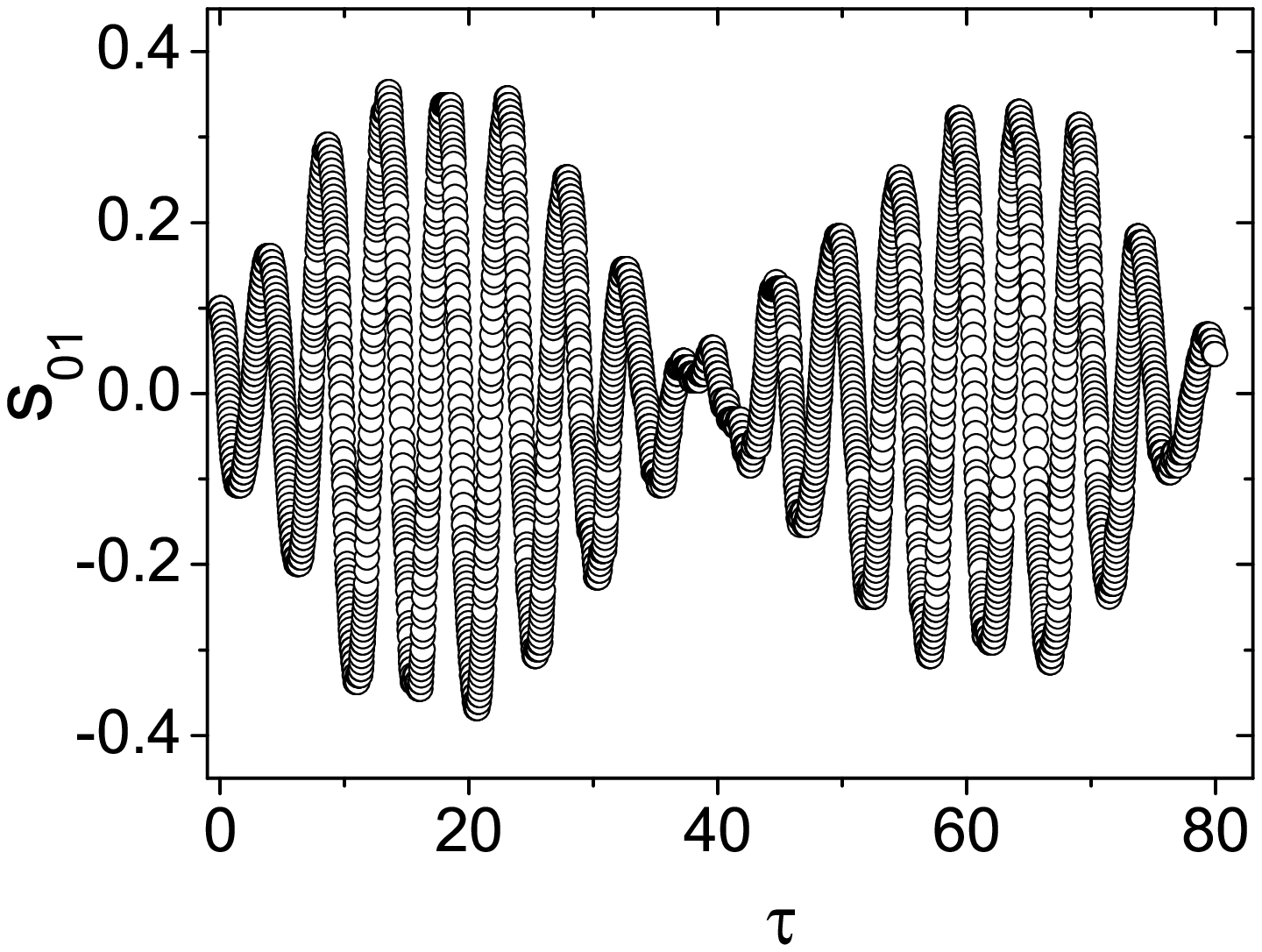}}
\caption{
Beating dynamics of displaced binary BEC solitons with an equal number of atoms $N_1=N_2=1$,  for $g_{12}=-0.2, s_{01}=0.1, s_{02}=1.3$ (left panel) and $g_{12}=-0.5, s_{01}=0.1, s_{02}=-1.3$ (right  panel).  Other parameters are fixed as $g_{11}^{(0)} =-1, g_{11}^{(1)}=-0.5, V_0=-0.5$.}
\label{figbeat}
\end{figure}

In Fig. \ref{fig3} we show typical dynamics of displaced binary solitons arising from a symmetric initial  displacements with respect to the effective potential minimum. The top two panels refers to  the case of  unequal numbers of atoms. We see that in this case the soliton components oscillate with the same frequency which depends on inter-species interaction strength and on number of atoms in the condensates (compare 1st and  2nd panel). For  $N_1\neq N_2$, however, the  oscillation frequencies of each component become unequal  (3rd panel) with appearance of  well-known beating phenomenon. The general solution in Eq. (\ref{gsolution}) shows that the beating phenomena  is expected also  for equal number of atoms and  small inter-species interactions if the motion is started with a generic initial displacement  $|s_{01}|\ne |s_{02}|$. This is exactly what the PDE calculations  in Fig. \ref{figbeat} show for the case $N_1=N_2=1$ and $|s_{01}|\ne |s_{02}|$ , in agreement with  our normal mode analysis.
\begin{figure}[htb!]
\centerline{
 \includegraphics[width=4cm,height=4cm]{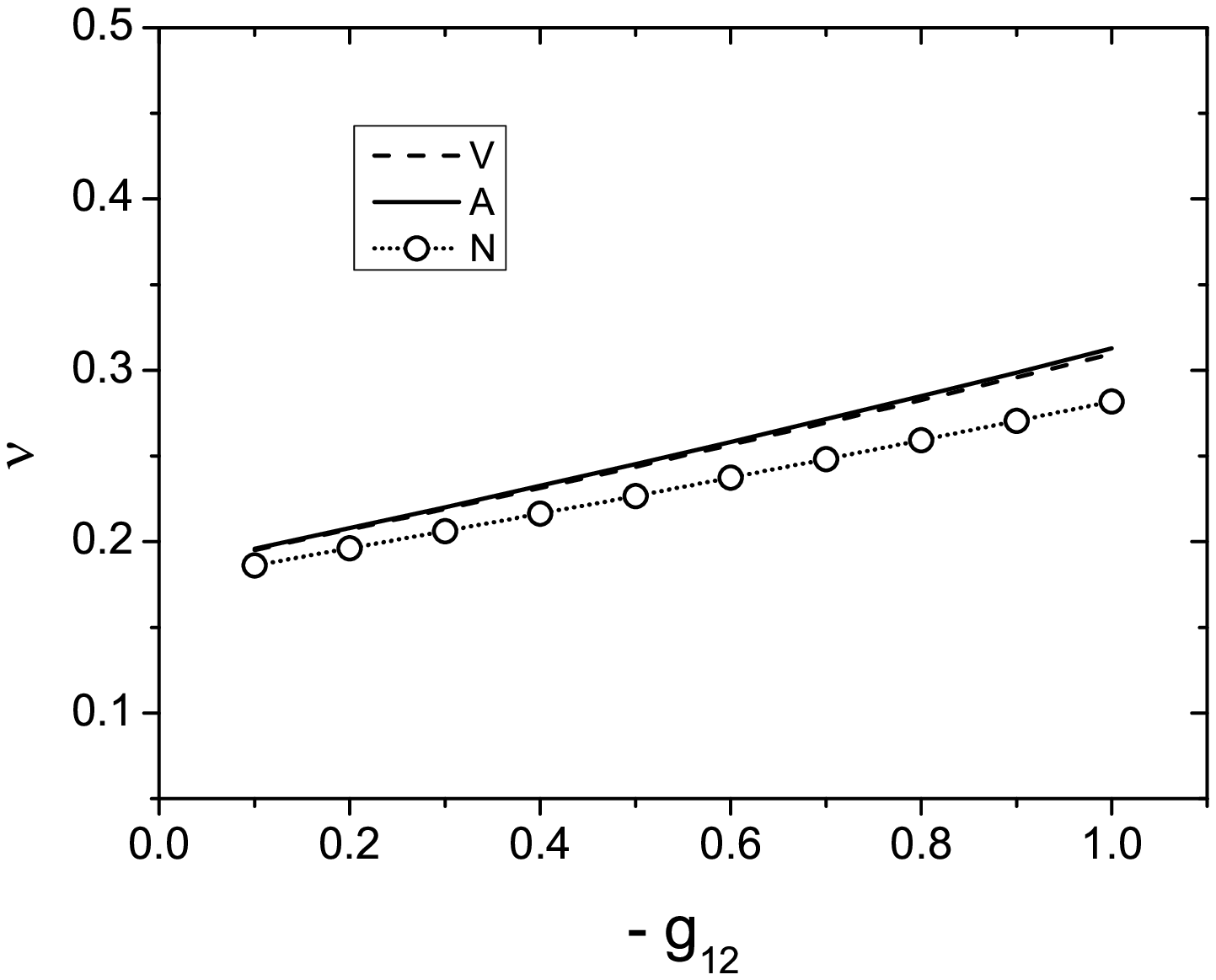}
\includegraphics[width=4cm,height=4cm]{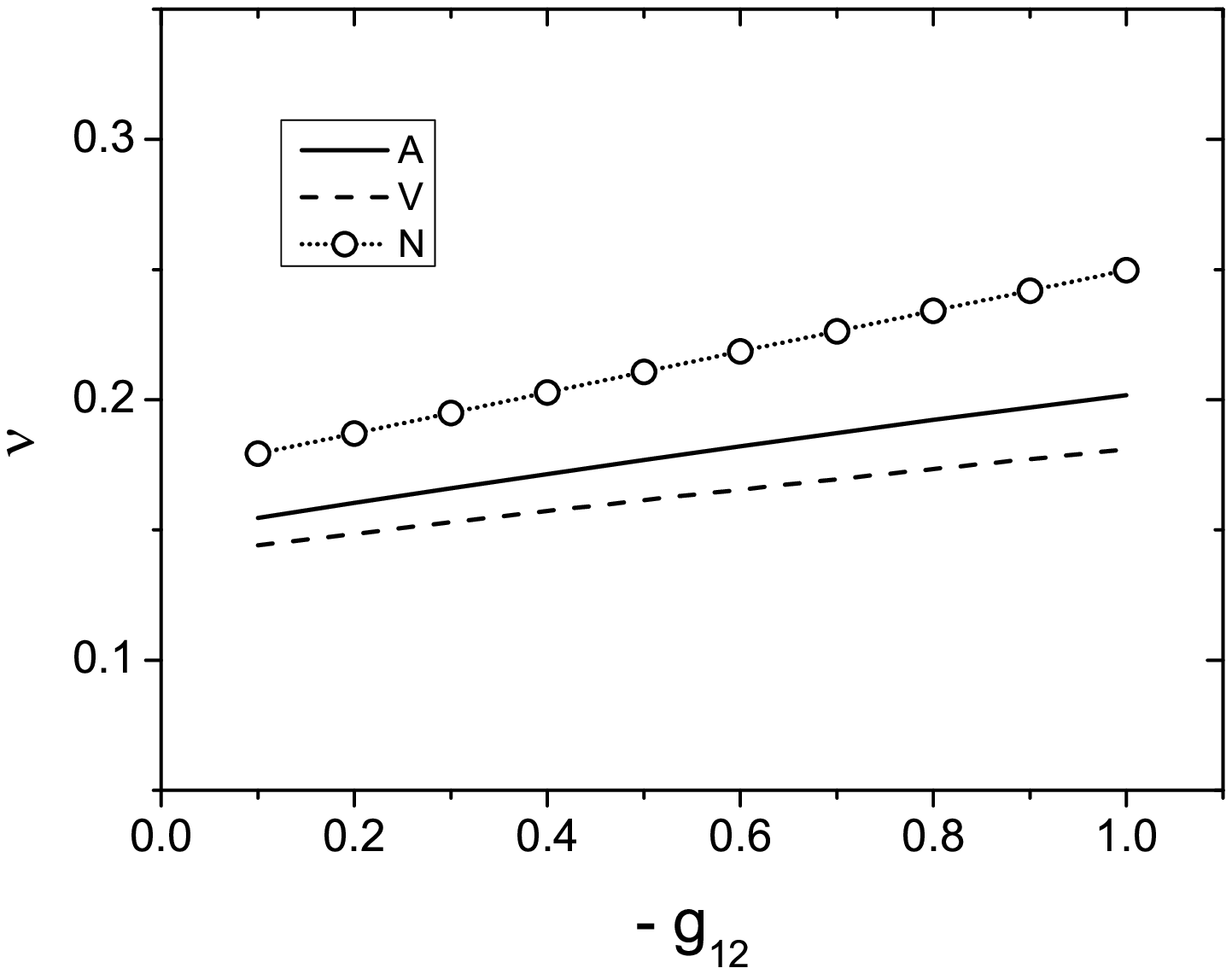}}
\centerline{
 \includegraphics[width=4cm,height=4cm]{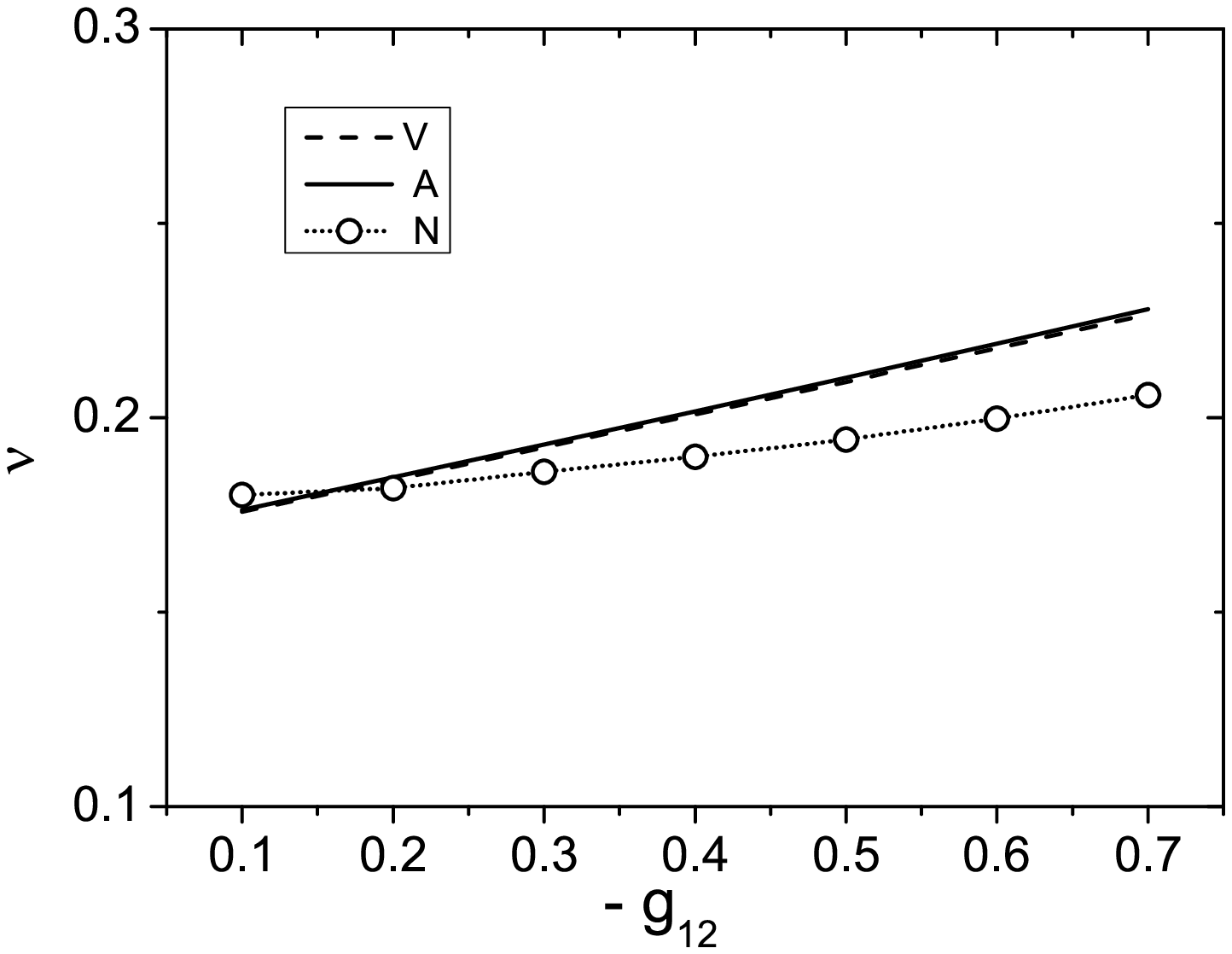}
\includegraphics[width=4cm,height=4cm]{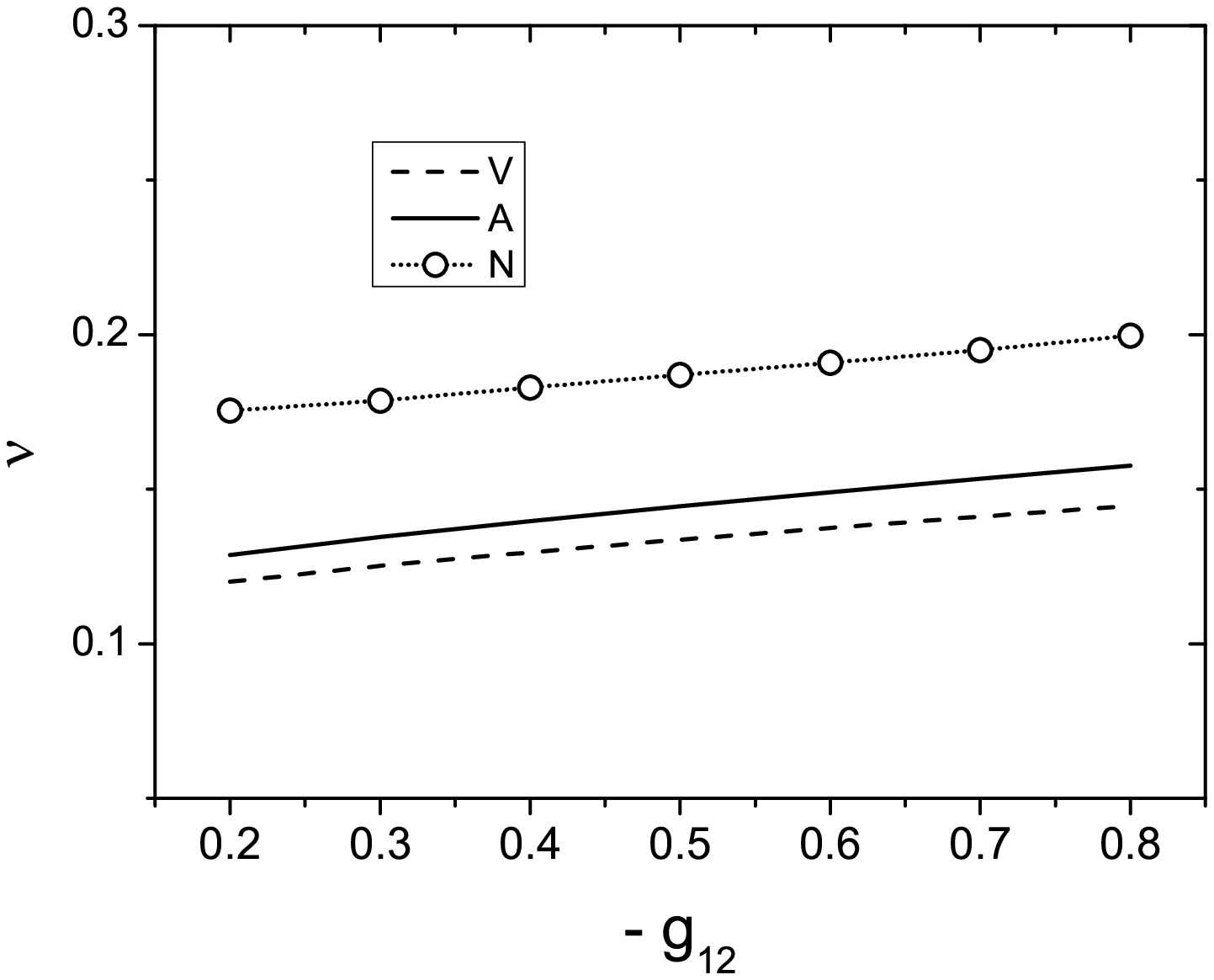}}
\centerline{
 \includegraphics[width=4cm,height=4cm]{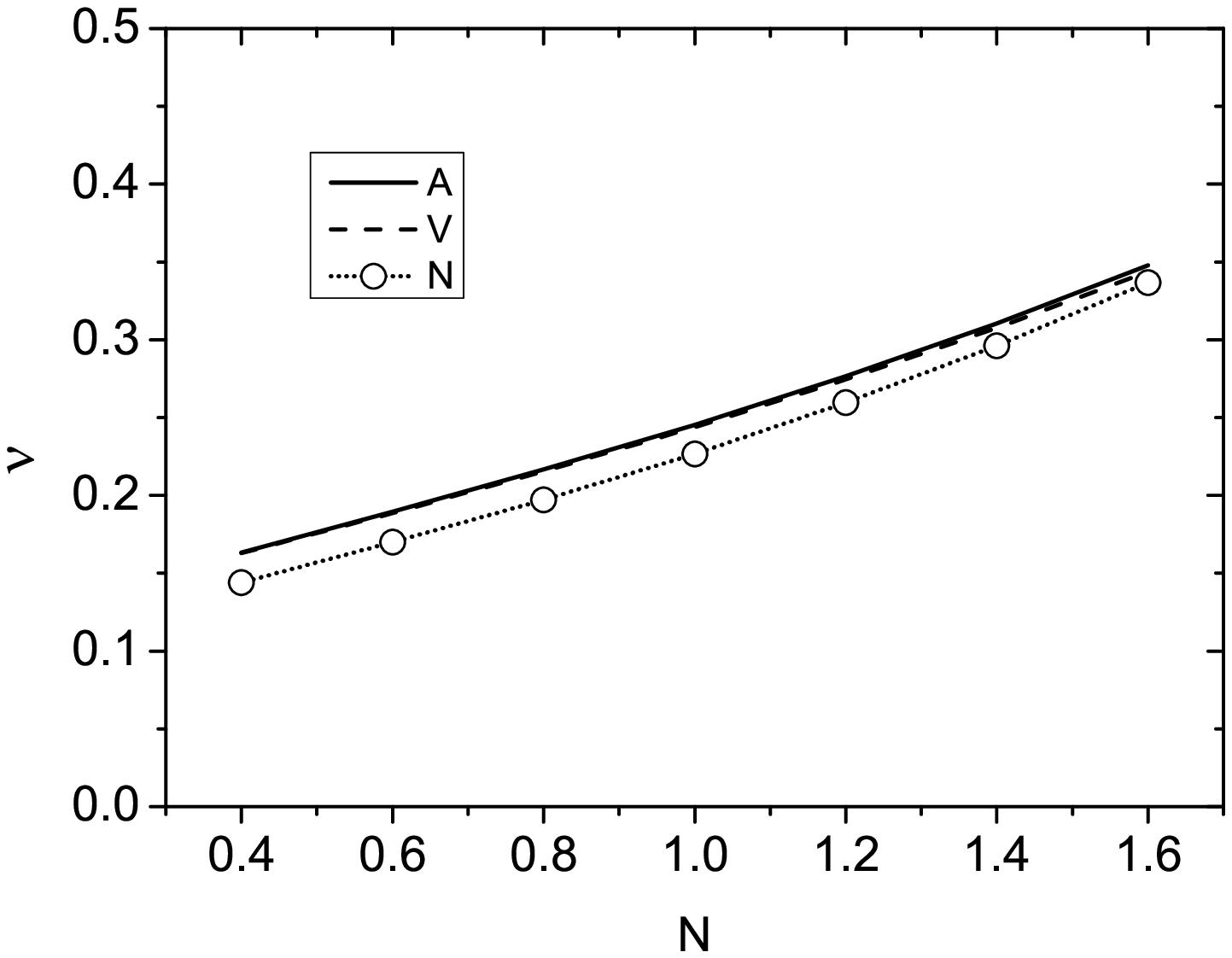}
\includegraphics[width=4cm,height=4cm]{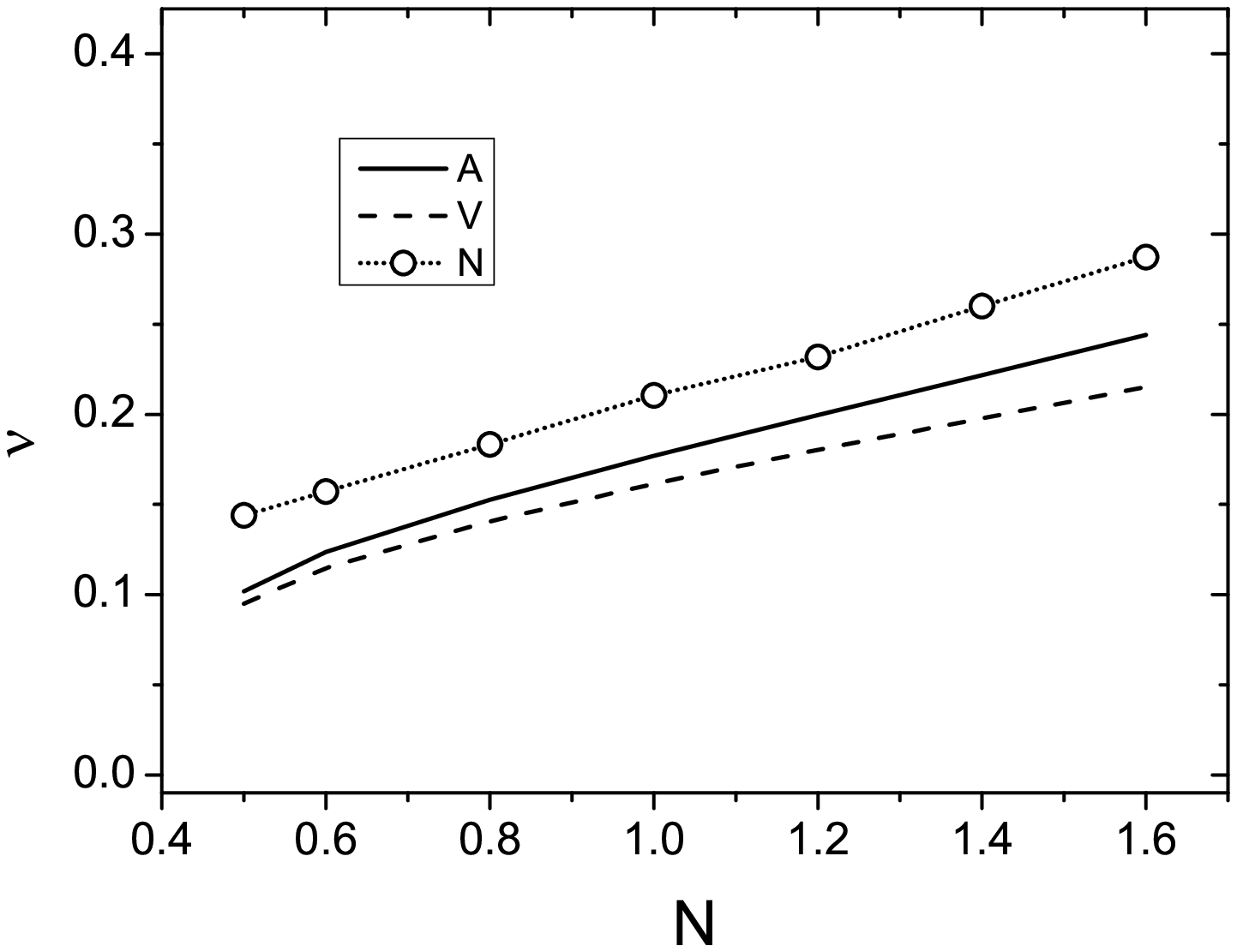}}
\caption{Top left panel. Frequency of the oscillation of the BEC components vs the inter-species interaction strength  for fixed number of atoms  $N_1=N_2=1$ and  $s_{0}=0.2$. Top right  panel: Same as that in the left panel but for a larger initial displacement $s_0=1.0$.
 Middle left panel. Frequency of the oscillation of the BEC components vs the inter-species interaction strength  for the case of unequal number of atoms  $N_1=1$ and $N_2=0.5$.  and $s_{0}=0.2$. Middle right panel: Same as that in corresponding left panel but for $s_0=1.0$.
 Bottom left panel. Frequency of the oscillation of BEC components  versus  the number of atoms for fixed inter-species interaction strength $g_{12} = -0.5$ and $s_0=0.2$.  Bottom right panel. Same as that in the left panel but $s_{0}=1.0$.
In all the panels the continuous curve represents the analytical expression in equation (\ref{eq15}), the dashed curve stands for result obtained from ODE in (\ref{eq13}) and open circles denote numerical GPE calculations. In each case, other parameters are fixed as: $ V_0 = -0.5,\, g_{11}^{(0)} = -1$ and $ g_{11}^{(1)} = -0.5$}
\label{fig4}
\end{figure}

The dependence of oscillation frequency ($\nu$) on $g_{12}$  is depicted  in Fig. \ref{fig4}  for the case $N_1/N_2=1$. In particular, the top left panel of this figure shows $\nu$ vs $-g_{12}$ for closely spaced solitons $(s_{0j}=\pm 0.1)$  while  the top right panel gives  the dependence on $g_{12}$  for $s_{0j}=\pm 0.5$. Notice that in both cases one can  estimate  values of $\nu$ very close to the exact numerical results (dotted curve)  from our simple analytical calculation. It may be pointed out that the variational approach gives  little higher values of $\nu$ than that of the numerical integration for $s_{0j}=\pm 0.1$( top panel) while we observed the opposite in the bottom panel. Also note that for overlapped condensates results obtained from analytical formula in $(\ref{eq15})$ and ODEs in ($\ref{eq13}$) are same. However, with the increase of separation deviation between these two results becomes appreciable. The deviation of analytical curve (solid) with ODE calculations (dashed) implies that the effects of aharmonicity has been captured. In addition, the curves in both the panels clearly indicate that the values of  $\nu$ increase as $g_{12}$  increases. This correlates with the fact that the corresponding effective potentials become more deep and acquire larger curvatures  at the origin as these parameters are increased, clearly leading to higher frequency values.

In the middle panel of Fig. \ref{fig4} we portrayed a plot similar to that in the top panel but with $N_1\neq N_2$ and observed same behavior of the frequency curve. However, the observed frequency in this case is little smaller than the previous one. This might  be associated with the fact that effective inter-species interaction is relatively larger $N_1=N_2$ that $N_1\neq N_2$ (Fig. \ref{fig1}).
%
\begin{figure}[htb]
\centerline{\includegraphics[width=4.5cm,height=4cm]{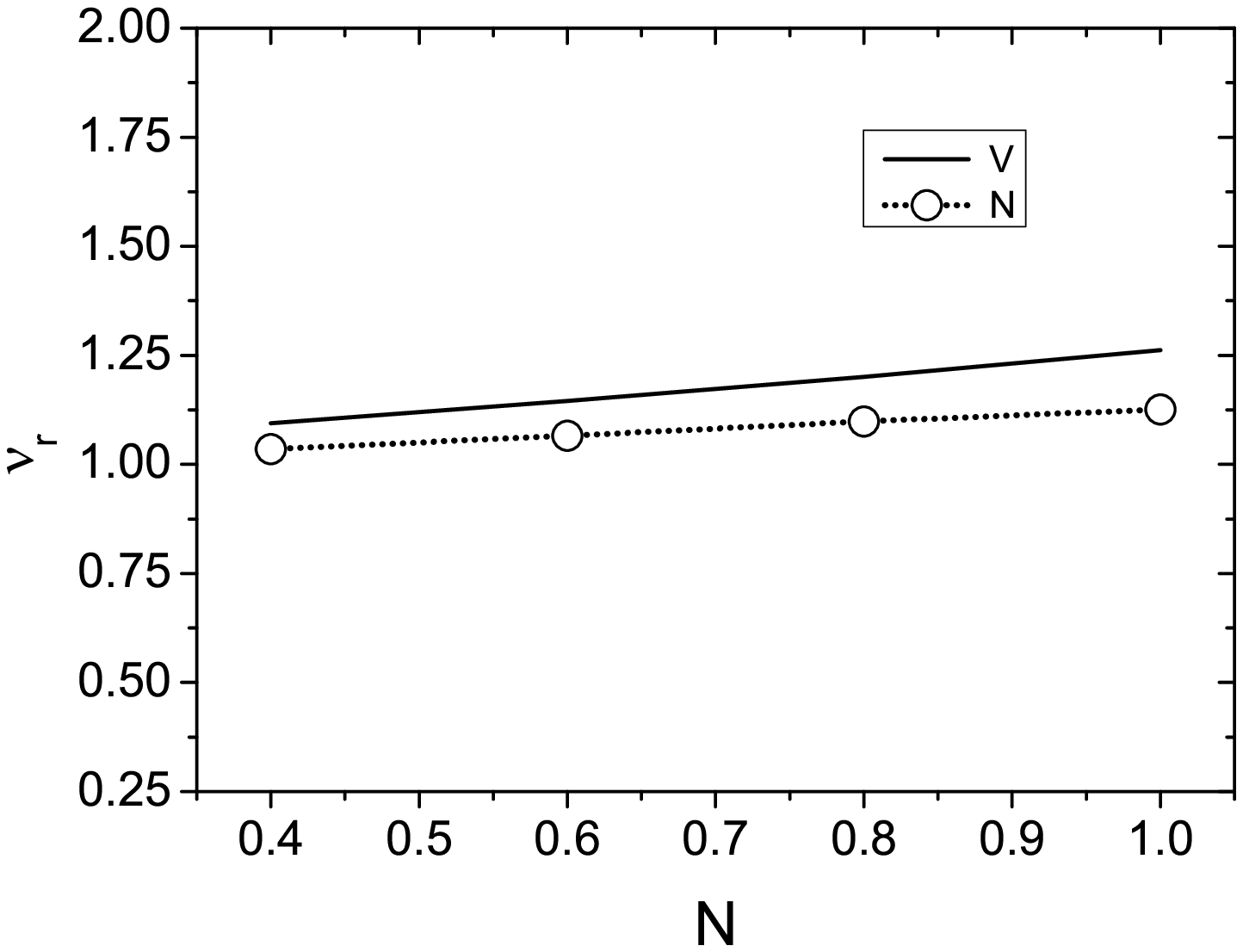}
\includegraphics[width=4.5cm,height=4cm]{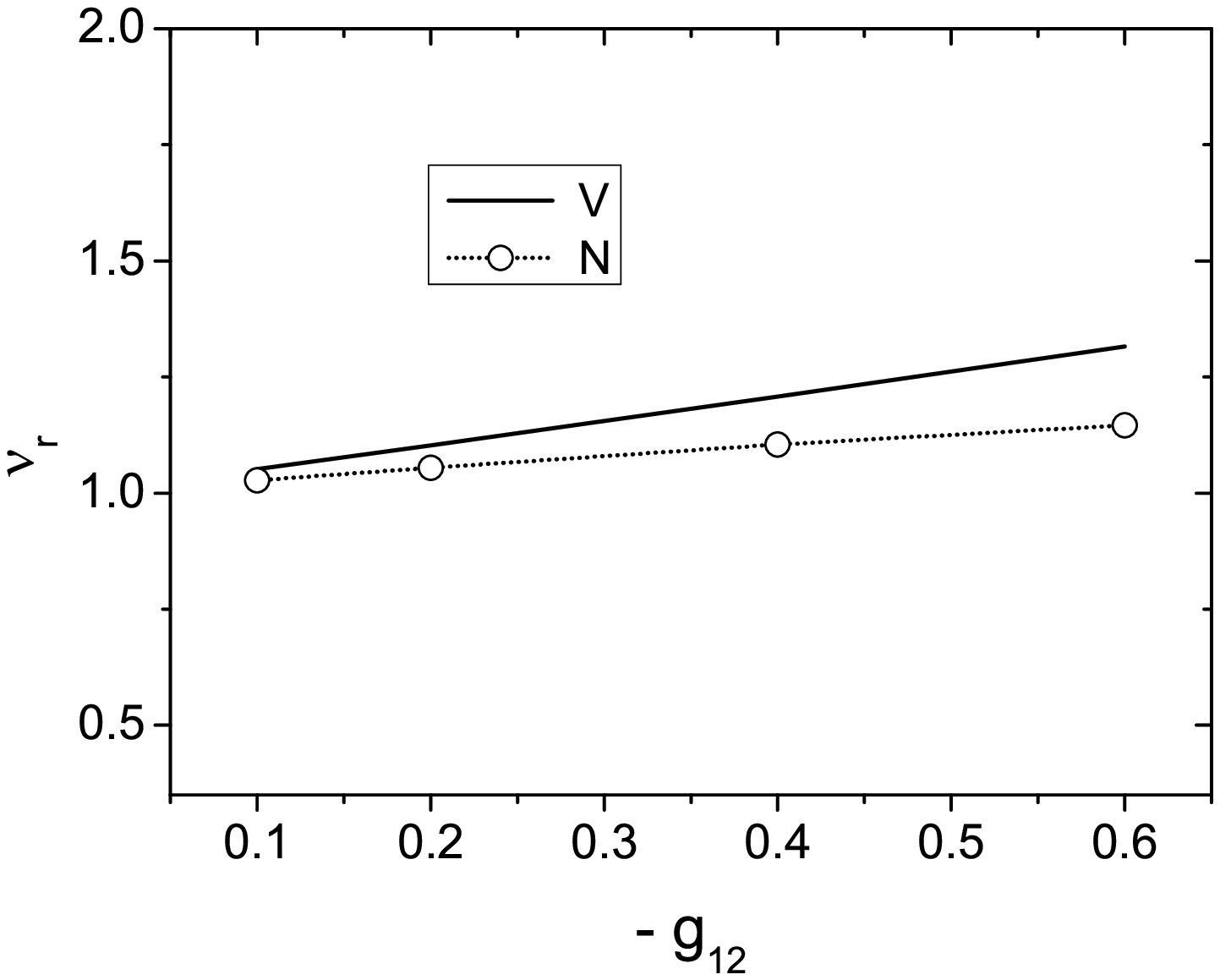}}
\vskip -0.5cm
\caption{
Relative frequency ($\nu_r=\frac{\omega_{asym}}{\omega_{sym}}$) of antisymmetric and symmetric modes vs number of atoms $N$ (left panel) and inter-species interaction $g_{12}$ (right panel). In both panels the solid curve refers to  Eq.(\ref{eqrfreq}) while the open dots refer to GPEs numerical integrations. Other parameters are fixed as $V_0=-0.5, g_{11}^{(0)}= -1, g_{11}^{(1)}=-0.5, s_0=0.2$.
}
\label{fig3a}
\end{figure}
In order to find dependence of  $\nu$ on number of atoms in the condensates, we consider  only the case $N=N_1=N_2$ and plot in the bottom  panels of Fig. \ref{fig4} variation of $\nu$ with $N$ for $g_{12}=-0.5$. Particularly, the left bottom  panel shows $\nu$ vs $N$ for $s_0=0.2$ while the right bottom panel gives a similar plot but $s_0=1.0$. As in the previous case, the frequency of oscillations here also increases with the increase of number of atoms due to the increase of effective inter-species interaction. Notice that the analytical (solid) and/or ODE (dashed) calculations here also gives a very good estimate of $\nu$ of the exact numerical calculation (dots with circles). The expected discrepancy of analytical curve from the ODE curve due to anharmonic effect is prominent for $s_0=1.0$ (bottom right panel).
In Sec. III we showed  that the frequency of the asymmetric mode is always greater than that of the symmetric one  with a ratio $\nu_r$ that  depends on both $g_{12}$ and $N$. In Fig. \ref{fig3a} we compare the dependence of  $\nu_r$  on $g_{12}$ and on $N$  as obtained from  Eq. (\ref{eqrfreq})  with the one obtained from PDE calculations. We see that in both cases a relatively good agreement is found.  It is also clear, both  from analytical and numerical results, that  $\nu_r$  increases with the increase of either inter-species interaction and number of atoms, and that  the  analytical results are slightly overestimating this growth.

\begin{figure}[htb]
\centerline{\includegraphics[width=3.1cm,height=3.75cm]{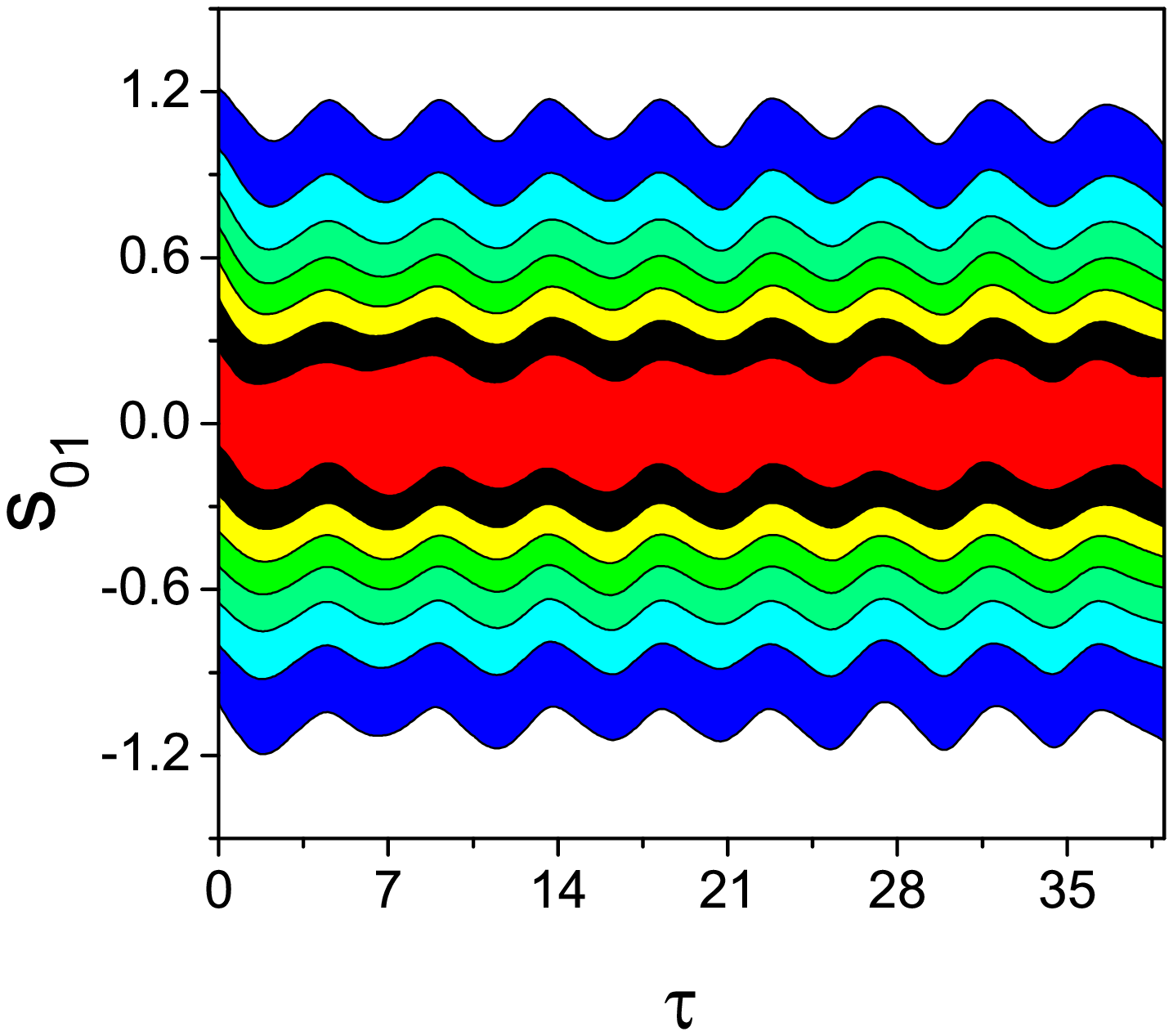}
\hskip -0.1cm
\includegraphics[width=3.1cm,height=3.75cm]{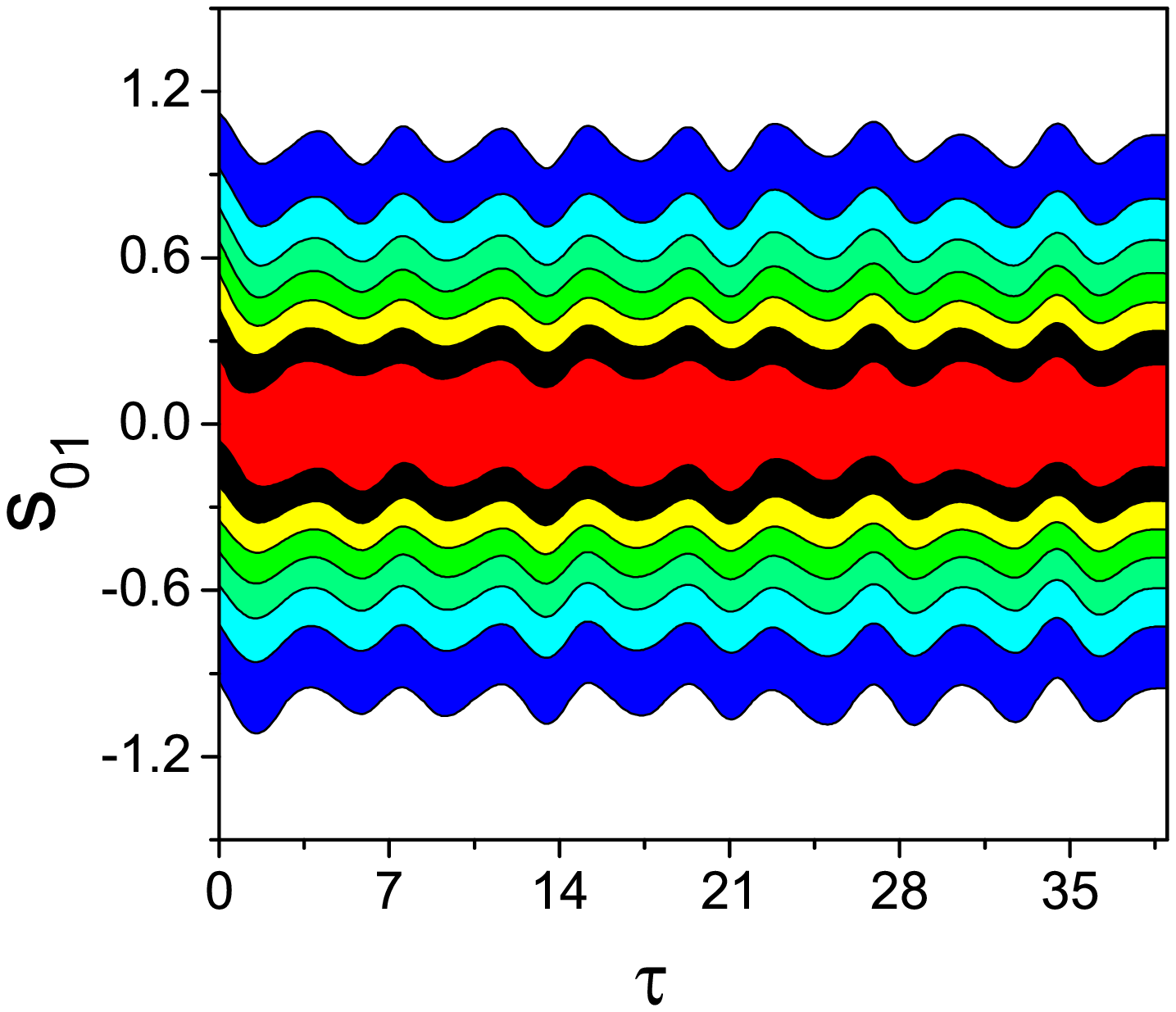}
\hskip-0.1cm
\includegraphics[width=3.1cm,height=3.75cm]{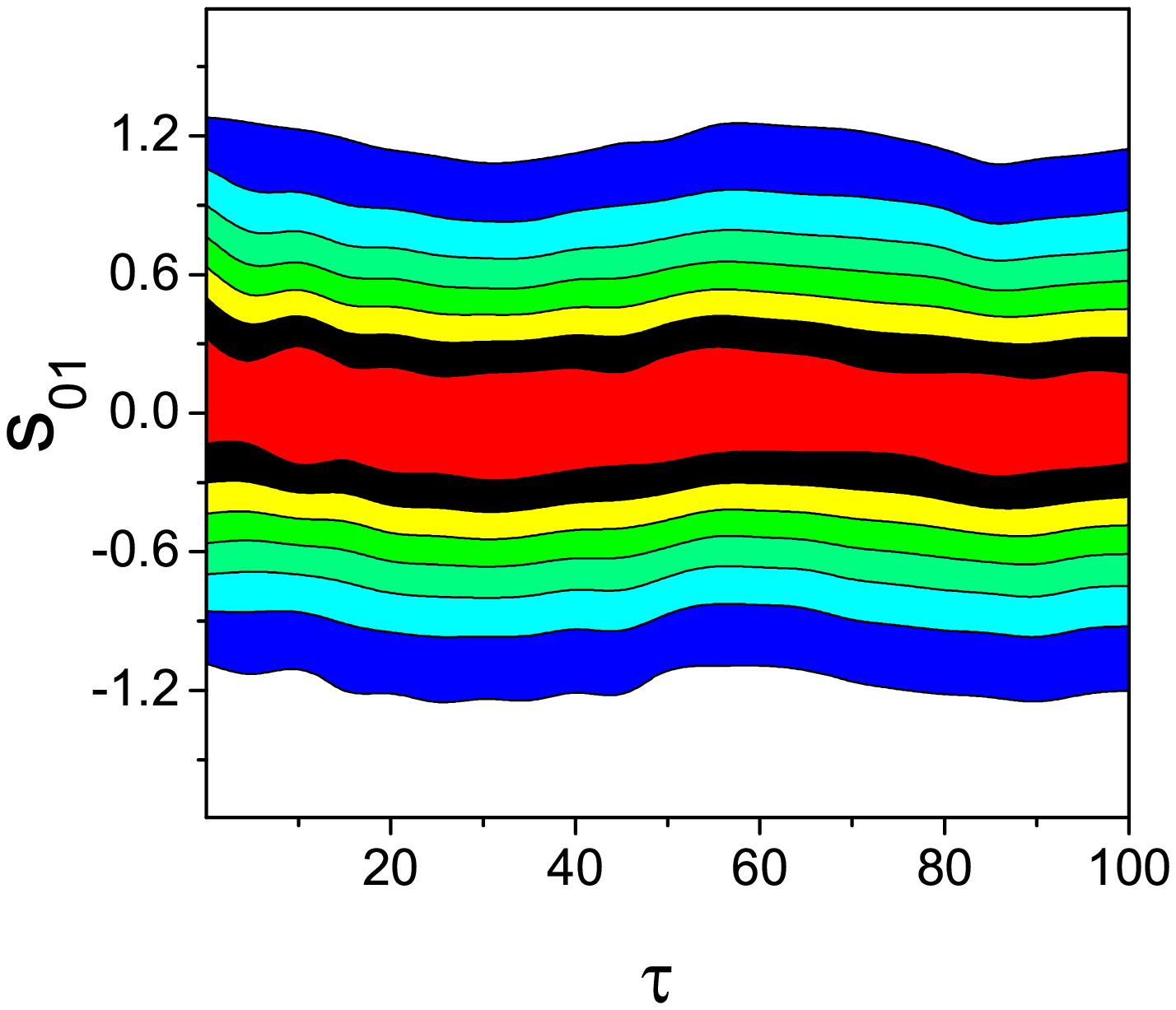}}
\vskip -0.1cm
\caption{Time evolution of displaced component binary soliton densities. In each contour plots we have taken slightly perturbed initial conditions for parameter values $ V_0 = -0.5,\, g_{11}^{(0)} = -1,\, g_{11}^{(1)} = -0.5$.  Left and middle panels show density evolution for $N_1=N_2=1$ and $g_{12}=-0.2$ (left panel) and for $N_1=N_2=0.8$ and $g_{12}=-0.5$. Right   panel  refers to the case of  unequal number of atoms $N_1=1, N_2=0.5$ for $g_{12}=-0.2$.
}
 \label{fig5}
\end{figure}

The stability of the oscillatory soliton dynamics has been checked with slightly perturbed initial BEC profiles  for  given values of parameters and then allowed them to evolve according the GPE copupled equations. Density plots for the evolution of soliton profiles for different values of $g_{12}$ and number of atoms is displayed in Fig. \ref{fig5}. This figure clearly indicates that during time evolution soliton profiles remain undistorted. To confirm the stabile evolution of the density profiles we slightly vary initial conditions and noticed that they still evolve uniformly with time.  The stability of soliton profile can also be examined from the phase plot of coupled ordinary differential equations in (\ref{eq11}). We have verified that in each case considered by us the phase plot exhibits stable focus.
%
\section{Conclusion}
  In this paper we have studied the dynamics of matter wave solitons of two-component Bose-Einstein condensates in combined linear and nonlinear optical lattices.

  In particular, we have investigated the dependence of  the oscillating dynamics resulting from two initially displaced  BEC soliton components on the inter-species interaction and on the number of atoms. 
  We showed that for small initial displacements binary solitons can be viewed as point masses connected by elastic springs of strengths related to  the amplitude  of the OL and to the intra and inter-species interactions. The displaced dynamics in can be decomposed in term of normal mode analysis from which analytical  expressions of the symmetric and anti-symmetric mode frequencies, are derived. The occurrence of beating phenomena both for unequal and for equal numbers of atoms for small interspecies interactions, was also predicted.
   The stability of the oscillating dynamics has been also investigated by direct numerical GPE integrations. The  predictions of the effective potential approach are found to be in quantitative agreement with direct numerical simulations. These  results suggest  the possibility to use dynamical behaviors of suitably prepared initial multi-component BEC solitons  as a tool for extracting  information about  physical characteristics of BEC mixtures such as interatomic interactions and species populations. In this respect, we remark that in contrast to intra-species interactions, direct measurements of the inter-species scattering lengths are more difficult to access. The  possibility to measure interspecies scattering lengths through  dynamical behaviors of displaced BEC components represents therefore an interesting  possibility which could be  tested  in real  experiments.
\\

\section*{Acknowledgment}
GAS wish to thank the Department of Physics of the University of Salerno for the hospitality received and for a one year research grant (AR$-2011$) during which this work has been done. MS acknowledges partial support from  the Ministero dell' Istruzione,
dell' Universit\'a e della Ricerca (MIUR) through a Programmi di Ricerca Scientifica di
Rilevante Interesse Nazionale (PRIN)-2008.

\end{document}